\documentclass[prd,superscriptaddress,preprintnumbers,twocolumn,showpacs,floatfix]{revtex4}
\usepackage{ifpdf}
\usepackage{graphicx}
\usepackage{amsmath}
\usepackage{amssymb}
\usepackage{color}
\usepackage{bm}

\begin{document}

\title{Bulk viscosity of spin-one color superconducting strange quark matter}

\author{Xinyang Wang}
\affiliation{Department of Physics, Arizona State University, Tempe, Arizona 85287, USA}

\author{Igor A. Shovkovy}
\email{igor.shovkovy@asu.edu}
\affiliation{Department of Physics, Arizona State University, Tempe, Arizona 85287, USA}
\affiliation{Department of Applied Sciences and Mathematics, Arizona State University, Mesa, Arizona 85212, USA}

\begin{abstract}
The bulk viscosity in spin-one color-superconducting strange quark matter is calculated 
by taking into account the interplay between the nonleptonic and semi-leptonic week 
processes. In agreement with previous studies, it is found that the inclusion of the 
semi-leptonic processes may result in non-negligible corrections to the bulk viscosity 
in a narrow window of temperatures. The effect is generally more pronounced for pulsars 
with longer periods. Compared to the normal phase, however, this effect due to the 
semi-leptonic processes is less pronounced in spin-one color superconductors. Assuming 
that the critical temperature of the phase transition is much larger than $40~\mbox{keV}$,
the main effect of spin-one color superconductivity in a wide range of temperatures is an 
overall increase of the bulk viscosity with respect to the normal phase. The corresponding 
enhancement factor reaches up to about 9 in the polar and {\it A}-phases, about 25 in the 
planar phase and about 29 in the CSL phase. This factor is determined by the suppression 
of the nonleptonic rate in color-superconducting matter and, therefore, may be even larger 
if all quark quasiparticles happen to be gapped. 
\end{abstract}

\date{September 2, 2010}

\pacs{12.38.Mh, 12.38.Aw, 12.15.Ji, 26.60.Dd}

\maketitle

\section{Introduction}

Understanding the physical properties of baryonic matter above nuclear saturation 
density is one of the fundamental challenges in modern nuclear astrophysics. Many 
aspects of neutron stars (e.g., the mass-radius relation, cooling and rotational 
dynamics, glitches and pulsar kicks) depend on these properties. For example, the 
equation of state of supranuclear baryonic matter plays the key role in determining 
the maximum possible mass of neutron stars. The harder (softer) equation of state is, 
the larger (smaller) maximum mass can be. The equation of state of dense baryonic 
matter is also one of the essential ingredients that determines the dynamics of 
core-collapse supernovae and, in turn, the mass distribution of black holes in the 
Universe \cite{core-collapse}. Several characteristics of the gravitational wave 
emission from merging neutron stars \cite{GW-emission}, which may be observed, e.g., 
by advanced Laser Interferometer Gravitational Wave Observatory (LIGO), are sensitive 
to the details of the equation of state. 

From the theoretical viewpoint, currently there is no consensus even regarding 
the qualitative state of matter at the highest densities reached in stellar cores. 
The most conservative possibility is that such matter is made of only nucleonic 
degrees of freedom. The study of several neutron stars in Ref.~\cite{Steiner:2010fz}, 
for example, does not exclude such a possibility, although a phase transition may be 
in agreement with their analysis, provided no extreme softening of the equation of 
state occurs. Another astrophysical determination of masses and radii of three neutron 
stars in Ref.~\cite{Ozel:2010fw} suggests, however, that the actual equation of state 
is too soft to be pure nucleonic. Such contradictory interpretations are representative
and show that the current knowledge is too limited to settle the issue. Theoretically, 
it may be also appropriate to mention that the observables associated with the equation 
of state alone have limited power to probe the actual nature of dense matter 
\cite{Ozel:2006bv,Alford:2006vz}. In fact, a true insight regarding the stellar interior 
may require a comprehensive understanding not only of the thermodynamical, but 
also transport properties and neutrino emission rates of various possible states of 
superdense matter. 

In this paper, we assume that baryonic matter at the highest stellar densities 
is deconfined quark matter. The possible formation of quark matter in stars is 
an old hypothesis \cite{quarkstars} that dates back to the time when the concept of quarks was 
first introduced \cite{GellMann:1964nj,Zweig:1981pd}. This is also supported by general considerations 
\cite{Collins:1974ky} based on the property of asymptotic freedom in quantum 
chromodynamics (QCD) \cite{asymptotic}. The main uncertainties of this scenario 
are (i) the value of the critical density, at which the deconfinement transition 
occurs, and (ii) the actual highest density reached in stars. If quark matter is 
formed, as we assume here, it is also likely to be a color-superconductor 
\cite{Barrois:1977xd,Bailin:1984dz}. (For reviews on color superconductivity 
and its general effects on stellar properties see, for example, 
Refs.~\cite{rev1,rev2,rev3,Buballa:2003qv,Shovkovy:2004me,Alford:2007xm}.) 
In this paper, in particular, we concentrate on the scenario, in which color 
superconductivity is due to same-flavor, spin-one Cooper pairing 
\cite{Schafer:2000tw,Schmitt:2002sc,Schmitt:2004et,spin-1}. 

The fact of liberation of quark degrees of freedom and the formation of a 
color-superconducting state of matter is likely to be revealed through a detailed 
study of the observational features of neutron stars. For example, one promising 
class of observables is related to the rates of weak processes. Such processes 
are known to be responsible for the cooling rates \cite{iwamoto} and damping 
of the rotational (r-mode) instabilities \cite{Andersson:1997xt} in stars. The
cooling is primarily determined by the neutrino emission rate, while the damping 
of r-modes is controlled by the viscosity of dense matter \cite{Madsen:1998qb}.

The bulk viscosity in the normal phase of three-flavor quark matter is usually dominated 
by the nonleptonic weak processes~\cite{WangLu,Sawyer,madsen,Xiaoping,Zheng,Dong:2007mb,Alford:2010gw}. 
It was argued in Ref.~\cite{Sa'd:2007ud}, however, that the interplay between the semi-leptonic 
and nonleptonic processes may be rather involved even in the normal phase of quark 
matter. Indeed, because of the resonance-like dynamics responsible for the bulk viscosity 
and because of a subtle interference between the two types of the weak processes, a 
larger rate of the nonleptonic processes may not automatically mean its dominant role.
In fact, it was shown that the contributions of the two types of weak processes are generally 
not separable and that, for a range of parameters, taking into account the semi-leptonic 
processes may substantially modify the nonleptonic result \cite{Sa'd:2007ud}. 

In this paper, we extend the analysis of Ref.~\cite{Sa'd:2007ud} and study the effect 
that spin-one color superconductivity has on the bulk viscosity when the interplay 
between the two types of weak processes is carefully taken into account. 
(For calculation of the bulk viscosity in other color superconducting phases see 
Refs.~\cite{Alford:2007rw,Alford:2006gy,Sa'd:2006qv,Sa'd:2008gf,Mannarelli:2009ia}.) 
The necessary ingredients for the calculation of the bulk viscosity are the rates of semi-leptonic 
(Urca) and nonleptonic weak processes. While the needed rates for the semi-leptonic 
processes in several spin-one color superconducting phases were obtained several 
years ago in Ref.~\cite{Sa'd:2006qv}, the corresponding rates of the nonleptonic 
processes remained unknown until very recently \cite{Wang:2009if}. Here we utilize 
both to obtain the bulk viscosity.

The rest of the paper is organized as follows. The general formalism for the calculation
of the bulk viscosity in strange quark matter with several active weak processes is 
reviewed in Sec.~\ref{Formalism}. This formalism is then used in Sec.~\ref{Calculation} 
to obtain our main results for the bulk viscosity as a function of temperature and the 
frequency of density oscillations. There we also study the enhancement effect of color 
superconductivity on the bulk viscosity, as well as the interplay of semi-leptonic and 
nonleptonic processes. In Sec.~\ref{Discussion}, we discuss the results and their 
potential implications for the physics of compact stars. Two appendices at the end of 
the paper contain our fits for the numerical suppression factors of the semi-leptonic 
and nonleptonic rates in spin-one color superconducting strange quark matter.

\section{Formalism}
\label{Formalism}

In this study, in order to calculate the bulk viscosity in the presence of several types of 
active weak processes, we follow the general formalism of Ref.~\cite{Sa'd:2007ud}. 
We assume that small oscillations of the quark matter density are described 
by $\delta n = \delta n_{0}\, \mbox{Re}(e^{i\omega t})$ where $\delta n_{0}$ is the 
magnitude of the oscillations. For such a periodic process, the bulk viscosity 
$\zeta$ is defined as the coefficient in the expression for the energy-density dissipation 
averaged over one period, $\tau=2 \pi/\omega$, 
\begin {equation}
\langle \dot{\cal E}_{\rm diss}\rangle =-\frac{\zeta}{\tau} 
\int_0^{\tau} dt \left(\nabla \cdot \vec v\right)^2,
\label{epsilon-kin}
\end{equation}
where $\vec v$ is the hydrodynamic velocity associated with the density oscillations. 
By making use of the continuity equation, $\dot{n}+n\,\nabla\cdot\vec v=0$, we derive
\begin{equation}
\langle{{\cal \dot{E}}_{\rm diss}}\rangle 
=-\frac{\zeta \omega^2}{2}\left(\frac {\delta n_0}{n}\right)^{2}.
\label{zeta-def}
\end{equation}
Such an energy-density dissipation of a pulsating hydrodynamic flow is the outcome 
of a net work done on a macroscopic volume over a period of the oscillation,
\begin{equation}
\langle \dot{\cal E}_{\rm diss}\rangle = \frac{n}{\tau} \int_0^{\tau} P \dot{\cal V} dt ,
\label{diss-energy}
\end{equation}
where ${\cal V}\equiv 1/n$ is the specific volume. 
By matching the hydrodynamic definition in Eq.~(\ref{zeta-def}) with the relation 
in Eq.~(\ref{diss-energy}), we derive the expression for the bulk viscosity,
\begin{equation}
\zeta=-\frac{2}{\omega^2}\left(\frac{n}{\delta n_0}\right)^2 \frac{n}{\tau}
       \int_0^{\tau} P \dot{\cal V} dt .
\label{zeta-pressure}
\end{equation}
The dominant mechanism behind the bulk viscosity is related to weak processes 
\cite{Madsen:1998qb,WangLu,Sawyer,madsen}. A periodic oscillation of the density 
is responsible for an instantaneous departure from $\beta$-equilibrium in the system. 
As a result, the forward and backward weak processes (e.g., $u+d\to s+u$ and 
$s+u \to u+d$), which have equal rates in equilibrium, become unbalanced. Their 
net effect is to restore the equilibrium composition.
However, since the weak rates are relatively slow, a substantial time lag between 
the oscillations of the fermion number density (and, thus, the specific volume) 
and the chemical composition (and, thus, the pressure) develops. If the resulting 
relative phase shift of the two oscillations is $\Delta \phi$, one finds from 
Eqs.~(\ref{diss-energy}) and (\ref{zeta-pressure}) that the corresponding energy 
dissipation and the bulk viscosity are proportional to $\sin\Delta \phi$. (Note 
that the departure from the thermal equilibrium is negligible because it is 
restored by strong forces on much shorter time scales.)

\begin{figure}[t]
\noindent
\includegraphics[width=0.35\textwidth]{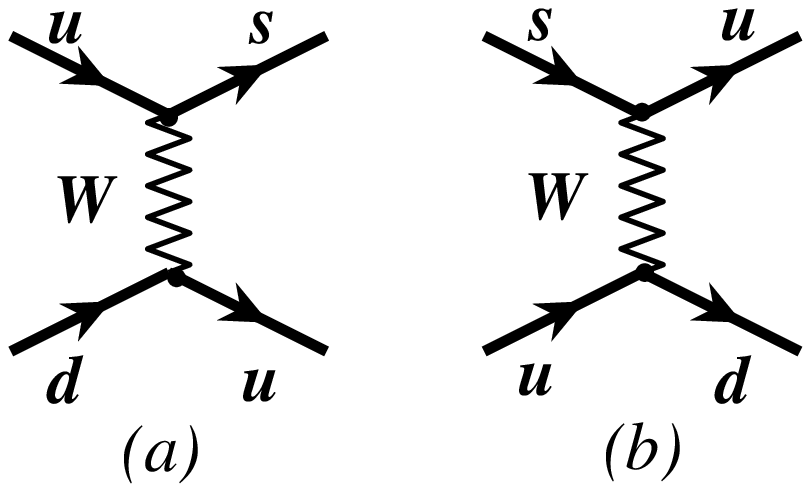}\\
\includegraphics[width=0.35\textwidth]{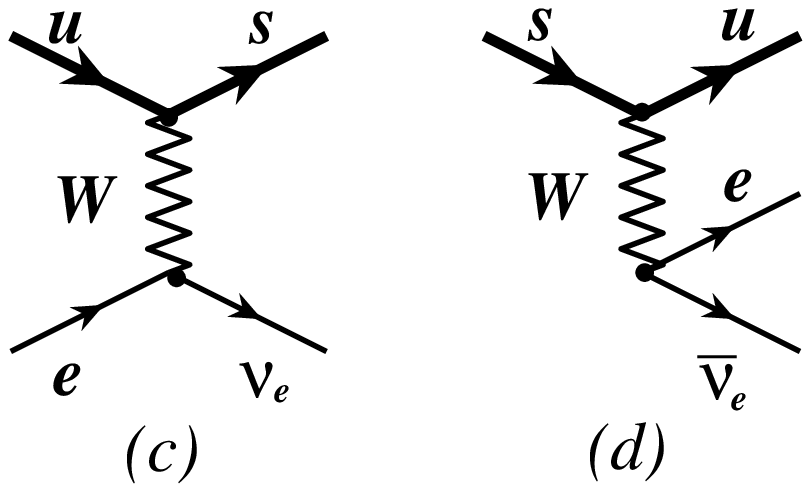}\\
\includegraphics[width=0.35\textwidth]{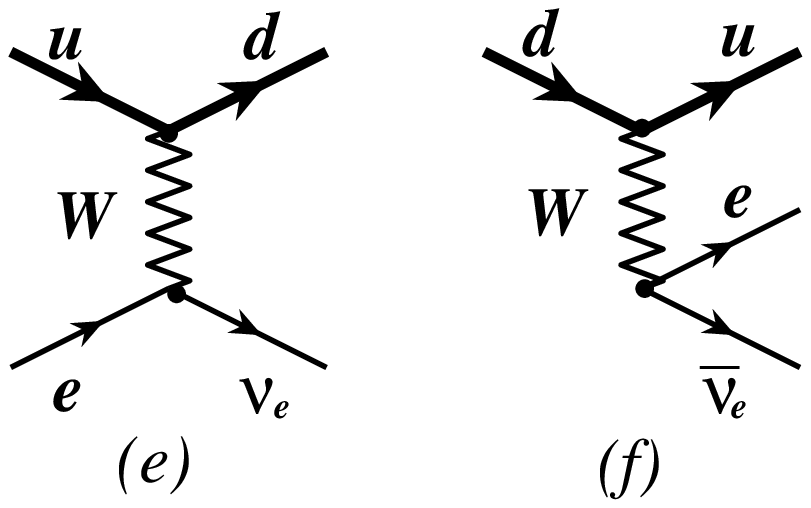}
\caption{Diagrammatic representation of the $\beta$ processes
that contribute to the bulk viscosity of dense quark matter.}
\label{fig-weak}
\end{figure}

It should be clear that the instantaneous flavor composition in oscillating 
quark matter and the rate difference of the forward and backward weak 
processes in Fig.~\ref{fig-weak} are related to each other. The difference 
of the rates changes the composition, while the composition in turn influences the 
difference of rates. The corresponding dynamics can be conveniently described 
in terms of the time dependent deviations of the chemical potentials from their 
equilibrium values. 

In $\beta$ equilibrium, the chemical potentials of the three lightest quarks 
are related as follows: $\mu_s=\mu_d=\mu_u+\mu_e$. Here $\mu_u$, $\mu_d$ and 
$\mu_s$ are the chemical potentials of up, down and strange quarks, while 
$\mu_e$ is the electron chemical potential. In pulsating matter, the 
instantaneous departure from equilibrium is described by the following two 
independent parameters:
\begin{subequations}
\begin{eqnarray}
\delta\mu_{1} &\equiv& \mu_s-\mu_d =\delta\mu_s-\delta\mu_d ,\\
\delta\mu_{2} &\equiv& \mu_s-\mu_u-\mu_e =\delta \mu_s-\delta \mu_u-\delta \mu_e,
\label{delta-mu-def}
\end{eqnarray}
\end{subequations}
where $\delta\mu_{i}$ denotes the deviation of the chemical potential $\mu_i$
from its equilibrium value. (Note that $\delta\mu_{3} \equiv \mu_d-\mu_u-\mu_e 
=\delta\mu_{2}-\delta\mu_{1}$ is not independent.) When $\delta\mu_{i}$ are 
non-zero, the corresponding pairs of forward and backward weak processes in 
Fig.~\ref{fig-weak} have different rates. To leading order, the rate differences 
are linear in $\delta\mu_{i}$, 
\begin{subequations}
\begin{eqnarray}
\Gamma_{(a)} - \Gamma_{(b)} &=& - \lambda_{1} \delta\mu_{1} ,
\label{ratediff1}\\
\Gamma_{(c)} - \Gamma_{(d)} &=& - \lambda_{2} \delta\mu_{2} ,
\label{ratediff2}\\
\Gamma_{(e)} - \Gamma_{(f)} &=& - \lambda_{3} \left(\delta\mu_{2}-\delta\mu_{1}\right) .
\label{ratediff3}
\end{eqnarray}
\label{ratediff0}
\end{subequations}
The corresponding $\lambda$-rates have been calculated for the normal phase \cite{WangLu,Sawyer,Madsen:1993xx,Anand} 
as well as several color superconducting phases of quark matter \cite{Alford:2006gy,Sa'd:2006qv,Wang:2009if}. 
The results for the normal phase, in particular, read
\begin{subequations}
\begin{eqnarray}
\lambda_{1}^{(0)} &\simeq& \frac{64}{5\pi^3} G_F^2
                     \cos^2 \theta_C \sin^2 \theta_C \mu_d^5 T^2 ,
\label{lambda1}\\
\lambda_{2}^{(0)} &\simeq& \frac{17}{40\pi} G_F^2
                     \sin^2 \theta_C \mu_s m_s^2 T^4 ,                 
\label{lambda2}\\
\lambda_{3}^{(0)} &\simeq& \frac{17}{15\pi^2} G_F^2
                     \cos^2 \theta_C \alpha _s \mu_d \mu_u \mu_e T^4 .
\label{lambda3}
\end{eqnarray}
\end{subequations}
These will be used below as a benchmark for the rates in spin-one color-superconducting phases. 

The semi-leptonic rate $\lambda_3  $ is determined by the Urca processes 
$u+e^-\to d+\nu_e$ and $d\to u+e^-+\bar{\nu}_e$, shown in diagrams $(e)$ and $(f)$ 
in Fig.~\ref{fig-weak}. It was calculated in Ref.~\cite{Sa'd:2006qv} for four 
different spin-one color-superconducting phases of quark matter. The result 
has a form of the product of the rate in the normal phase $\lambda^{(0)}_3$ and 
a phase-specific suppression factor, 
\begin{equation}
\lambda_3  = \lambda^{(0)}_3\left[\frac13 +\frac23 H\left(\frac{\phi}{T}\right)\right],
\label{lambda_3_general}
\end{equation}
where $\phi$ is the spin-one color-superconducting gap parameter, and 
$H(\phi/T)$ is a suppression factor for the processes involving gapped quasiparticles.
(The first term in square brackets is the contribution of ungapped quasiparticles.)
When $\phi \to 0$, the suppression factor $H(\phi/T)$ approaches $1$ and the normal 
phase result is restored. A simple fit to the numerical data of Ref.~\cite{Sa'd:2006qv} 
for $H(\phi/T)$ is presented in Appendix~\ref{Urca-rates}.

Because of similar kinematics and phase space constraints for the other pair of semi-leptonic 
processes, $u+e^-\to s+\nu_e$ and $s\to u+e^-+\bar{\nu}_e$, shown in diagrams $(c)$ and $(d)$ 
in Fig.~\ref{fig-weak}, the dependence of the rate $\lambda_2$ on the color-superconducting 
gap should take the same form as $\lambda_3$ in Eq.~(\ref{lambda_3_general}), i.e.,
\begin{equation}
\lambda_2  = \lambda^{(0)}_2\left[\frac13 +\frac23 H\left(\frac{\phi}{T}\right)\right].
\label{lambda_2_general}
\end{equation}
In contrast, the rate $\lambda_1 $ is determined by the nonleptonic processes $u+d\to s+u$ 
and $s+u\to u+d$, see diagrams $(a)$ and $(b)$ in Fig.~\ref{fig-weak}, which have a 
qualitatively different kinematics. In spin-one color-superconducting phases of quark matter, 
this was recently calculated in Ref.~\cite{Wang:2009if}. The numerical result can be conveniently 
summarized by the following expression:
\begin{equation}
\lambda_1 = \lambda^{(0)}_1  \left[{\cal N} + (1-{\cal N})\tilde{H}\left(\frac{\phi}{T}\right)\right],
\label{lambda_1_general}
\end{equation}
where, in addition to the suppression factor $\tilde{H}(\phi/T)$, we also introduced a constant 
${\cal N}$, which determines a relative contribution of the ungapped quasiparticles to the 
corresponding rate. In the four spin-one phases studied in Ref.~\cite{Wang:2009if}, the constant 
takes the following values: ${\cal N}^{A} = {\cal N}^{\rm polar} = 1/9$, ${\cal N}^{\rm planar} 
\approx 0.0393$, and ${\cal N}^{\rm CSL} = 928/27027\approx 0.0343$. A simple fit to the numerical 
data for $\tilde{H}(\phi/T)$ is given in Appendix~\ref{nonleptonic-rates}. 

When the rates (\ref{lambda_3_general}), (\ref{lambda_2_general}) and (\ref{lambda_1_general}) are 
known, the calculation of the instantaneous pressure and, thus, the bulk viscosity from Eq.~(\ref{zeta-pressure}) 
is straightforward \cite{Sa'd:2007ud}. Here we quote only the final expression for the viscosity,
\begin{equation}
\zeta=\zeta_{1}+\zeta_{2}+\zeta_{3},
\label{zeta-general}
\end{equation}
where
\begin{subequations}
\begin{eqnarray}
\zeta_{1} &=& \frac{n}{\omega}\frac{\alpha_{2}\alpha_{3}}{g_{1}^2+g_{2}^2}
\Big[\alpha_{1}\alpha_{2}\alpha_{3} C_{1}^2   \nonumber\\
&+&\left(\alpha_{1}+\alpha_{2}+\alpha_{3}\right)
\left(A_{1}C_{2}-A_{2}C_{1}\right)^2\Big],
\label{zeta11} \\
\zeta_{2} &=& \frac{n}{\omega}\frac{\alpha_{1}\alpha_{3}}{g_{1}^2+g_{2}^2}
\Big[
\alpha_{1}\alpha_{2}\alpha_{3} C_{2}^2\nonumber\\
&+&\left(\alpha_{1}+\alpha_{2}+\alpha_{3}\right)
\left[\left(A_{2}-B_{2} \right) C_{1} -A_{2}C_{2}\right]^2\Big],
\label{zeta22} \\
\zeta_{3} &=& \frac{n}{\omega}\frac{\alpha_{1}\alpha_{2}}{g_{1}^2+g_{2}^2}
\Big[
\alpha_{1}\alpha_{2}\alpha_{3} \left(C_{1}-C_{2}\right)^2\nonumber\\
&+&\left(\alpha_{1}+\alpha_{2}+\alpha_{3}\right)
\left(B_{1}C_{2}-B_{2}C_{1}\right)^2\Big],
\label{zeta33}
\end{eqnarray}
\label{zeta1-3}
\end{subequations}
and
\begin{subequations}
\begin{eqnarray}
 \hspace{-10mm}&&
 g_{1} = -\alpha_{1}\alpha_{2}\alpha_{3}
+ \left(\alpha_{1} + \alpha_{2} + \alpha_{3} \right)
\left(B_{1} A_{2}-A_{1} B_{2}\right),\\
 \hspace{-10mm}&&
 g_{2}  =  \alpha_{1}\alpha_{2} \left(B_{1} - B_{2}\right)
+\alpha_{1}\alpha_{3}\left(A_{2}-B_{2}\right)
+\alpha_{2}\alpha_{3} A_{1}.
\end{eqnarray}
\end{subequations}
Here $\alpha_{i}\equiv n \omega/\lambda_{i}$ ($i=1,2$) and $n$ is the baryon density of 
quark matter. The quantities $A_i$, $B_i$ and $C_i$ are susceptibility-like functions, 
see Ref.~\cite{Sa'd:2007ud} for the definition. To leading order in $\phi/\mu_i$, they 
are the same as in the normal phase. 

For comparison, let us also note that the bulk viscosity in the limit of the vanishing 
semi-leptonic rates reads
\begin{equation}
\zeta_{\rm non} =\frac{n}{\omega}\frac{\alpha_1 C_1^2}{\alpha_1^2+A_1^2}.
\label{zeta_non}
\end{equation}

\section{Numerical results for bulk viscosity}
\label{Calculation} 

In our calculation of the bulk viscosity in spin-one color-superconducting quark matter below, 
we choose the same two representative sets of model parameters as in Ref.~\cite{Sa'd:2007ud}:
\begin{table}[h]
\label{TableAB}
\begin{ruledtabular}
\begin{tabular}{ll}
Set A & Set B\\
\hline
$n = 5\rho_0$ & $n = 10\rho_0$\\
$m_s = 300$~MeV & $m_s = 140$~MeV\\
$\alpha_s = 0.2$ & $\alpha_s  = 0.1$\\
\end{tabular}
\end{ruledtabular}
\end{table}

In both cases, the masses of light quarks are the same: $m_u =  5$~MeV and $m_d =  9$~MeV.
In accordance with general expectations, the values of the strange quark mass $m_s$
and the strong coupling constant $\alpha_s$ should be larger (smaller) in the case of 
lower (higher) density. This qualitative property is reflected in the model parameters 
in Set A (Set B). The values of all chemical potentials as well as the coefficient functions 
$A_{i}$, $B_{i}$ and $C_{i}$ for each set of parameters are quoted in Table~\ref{TableI}. 

It may be appropriate to briefly comment about the choice of the strong coupling constant
$\alpha_s$ in the model at hand. The values of $\alpha_s$ in both sets of parameters may 
seem abnormally small. Indeed, the running coupling in QCD is about $0.12$ at the scale of 
$M_Z$ (mass of $Z$ boson) and about $0.32$ at $\sqrt{3}~\mbox{GeV}$ \cite{Ellis:1996xc}. 
However, here we use the model parameter $\alpha_s$ only in order to capture several 
qualitative (Fermi liquid) effects in quark matter. Its nonzero value allows (i) to avoid the 
underestimation of the rate of semi-leptonic processes due to a limited phase space \cite{iwamoto} 
and (ii) to mimic the modification of the quark equation of state due to strong interactions, 
see Ref.~\cite{Sa'd:2007ud} for details. The naive extension of the corresponding leading 
order corrections to the regime of strong coupling is problematic. Not only this would imply 
the use of the perturbative results beyond the range of their validity, but this would also lead 
to very large and seemingly unphysical effects on the equation of state, used to determine 
the susceptibility functions $A_i$, $B_i$ and $C_i$. (Notably, if the equation of state is
kept unchanged, the increase of $\alpha_s$ in the $\lambda_{3}^{(0)}$-rate, even by an 
order of magnitude, has little effect on the viscosity.) This dilemma could be resolved by 
properly accounting the non-perturbative dynamics of QCD. At present, however, such a task 
seems insurmountable at the low energy scales relevant for neutron stars. For the purposes 
of this study, therefore, we treat $\alpha_s$ as a small independent parameter that captures 
only some qualitative properties of quark matter. 

Here the critical temperature of the spin-one color-superconducting phase transition is 
assumed to be $T_c=2~\mbox{MeV}$. This may be a somewhat high, but still reasonable 
value for $T_c$. Indeed, in QCD the spin-one gap is estimated to be about two orders of 
magnitude smaller than the spin-zero gap \cite{Schafer:2000tw,Schmitt:2002sc,Schmitt:2004et}, 
and the latter is naturally of order $100~\mbox{MeV}$ 
\cite{rev1,rev2,rev3,Buballa:2003qv,Shovkovy:2004me,Alford:2007xm}. Even higher 
values of the spin-one gap have been reported in Ref.~\cite{Marhauser:2006hy}. 
The effect of varying the critical temperature is easy to understand and will be briefly 
discussed below. As in Ref.~\cite{Sa'd:2006qv}, we use the following model temperature 
dependence of the gap parameter:
\begin{equation}
\phi(T)=\phi_{0}\,\sqrt{1-\left(\frac{T}{T_c}\right)^2},\quad \mbox{for}\quad T<T_c
\label{Phi_of_T}
\end{equation}
with $\phi_{0}$ being the value of the gap parameter at $T=0$. Note that the ratio 
$T_c/\phi_{0}$ depends on the choice of the phase \cite{Schmitt:2004et}.
The approximate values of this ratio are $0.8$ (CSL), $0.66$ (planar),
$0.49$ (polar), and $0.81$ ({\it A}-phase).

\begin{table*}
\caption{Two sets of parameters used in the calculation of the bulk viscosity. }
\label{TableI}
\begin{ruledtabular}
\begin{tabular}{ccccccccccc}
model   & $\mu_e$ {[MeV]}  & $\mu_u$ {[MeV]}  & $\mu_d=\mu_s$ {[MeV]}   
& $A_1$ {[MeV]} & $A_2$ {[MeV]} 
& $B_1$ {[MeV]} & $B_2$ {[MeV]} 
& $C_1$ {[MeV]} & $C_2$  {[MeV]} \\
\hline
Set A &  $39.139$  & $402.463$  & $441.602$ 
& $239.432$ & $127.937$ 
& $111.386$ & $-3.726\times 10^4$ 
& $-60.463$ & $-60.460$ \\
Set B &  $7.396$  & $495.275$  & $502.671$ 
& $324.556$ & $164.288$ 
& $160.268$ & $-2.080\times 10^6$ 
& $-10.692$ & $-10.709$ \\
\end{tabular}
\end{ruledtabular}
\end{table*}

\begin{figure*}[t]
\includegraphics[width=0.48\textwidth]{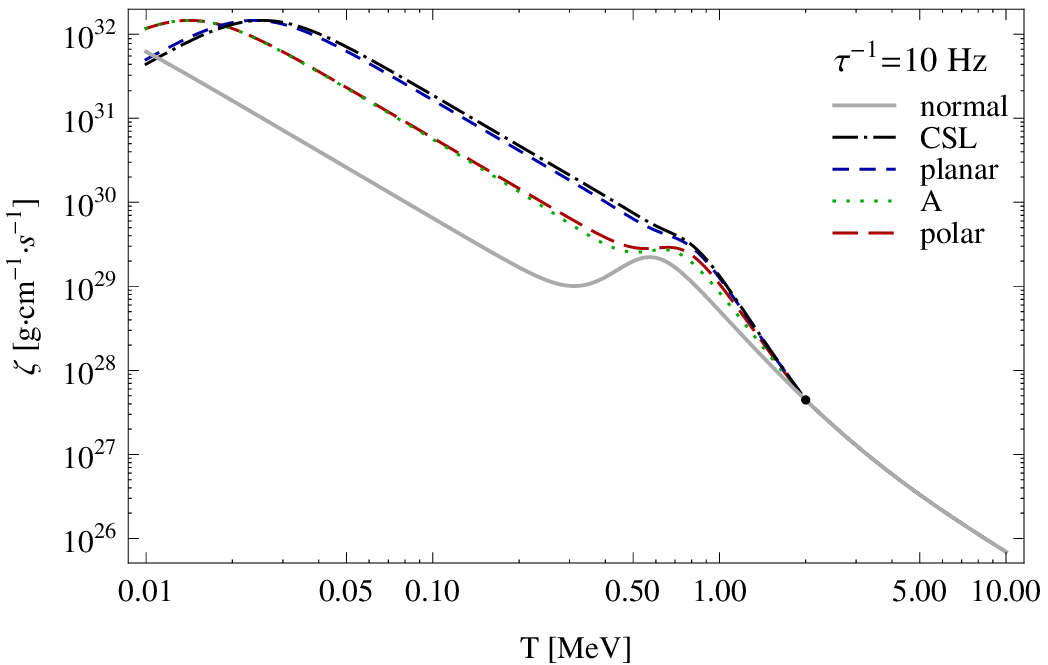}\hspace{0.03\textwidth}
\includegraphics[width=0.48\textwidth]{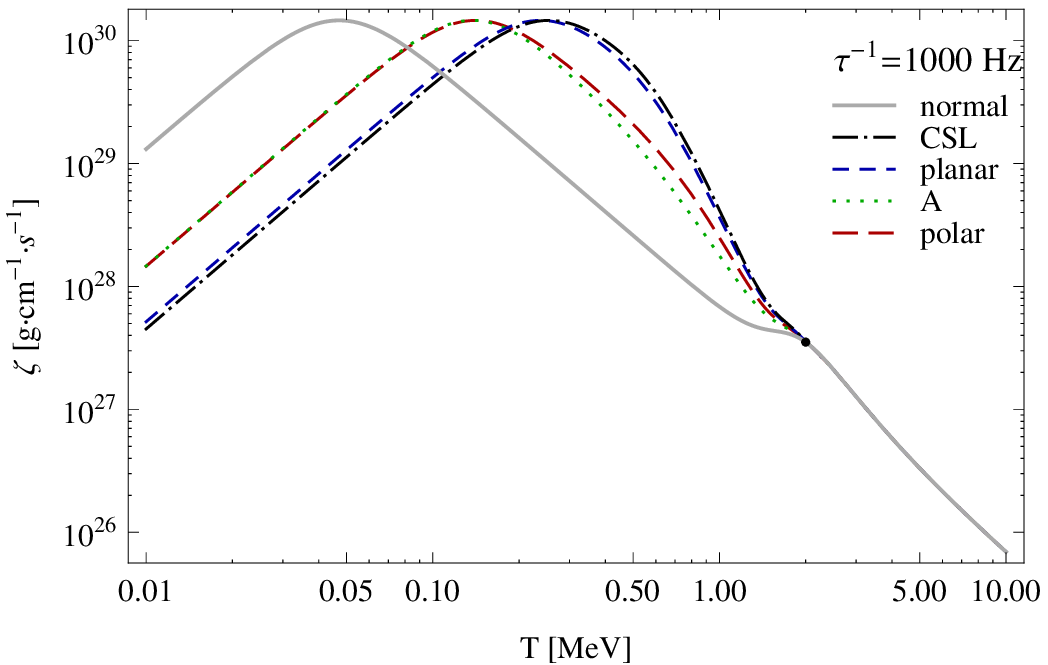}\\
\includegraphics[width=0.48\textwidth]{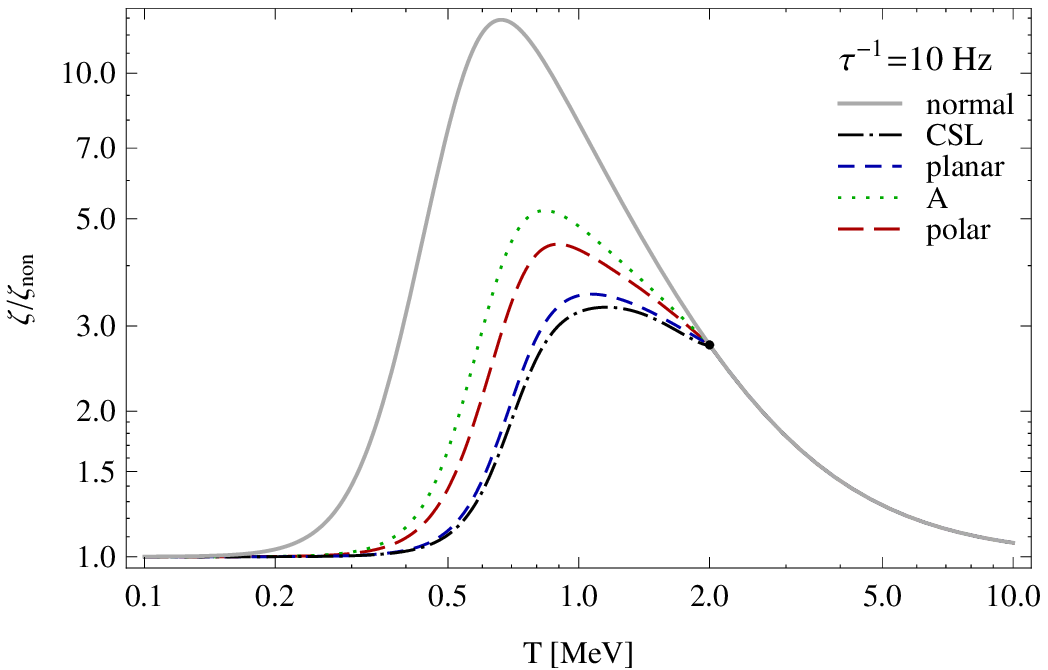}\hspace{0.03\textwidth}
\includegraphics[width=0.48\textwidth]{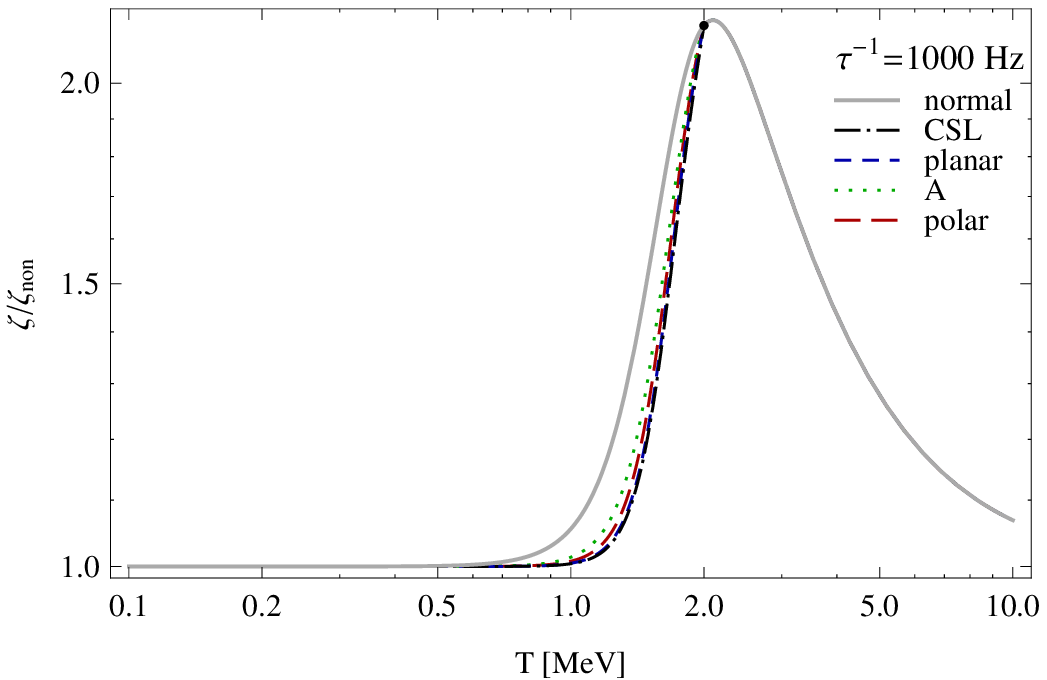}
\caption{(color online). Temperature dependence of bulk viscosity $\zeta$ and the ratio $\zeta/\zeta_{\rm non}$  
for model parameters in Set A and the spin-one color-superconducting critical temperature $T_c=2~\mbox{MeV}$. 
The results for two fixed frequencies of the density oscillations are shown.}
\label{results-set-A}
\end{figure*} 

For model parameters in Set A, the numerical results are  presented in Fig.~\ref{results-set-A}. 
As we can see, the value of $T_c$ determines the point where the bulk viscosity starts to 
deviate from the benchmark result in the normal phase (shown by the gray solid line). The upper 
panels show the dependence of the bulk viscosity $\zeta$ on temperature for two representative 
values of the oscillation frequency, $\tau^{-1}=10~\mbox{Hz}$ and $\tau^{-1}=1000~\mbox{Hz}$. 
The lower panels in the same figure show the temperature dependence of the ratio 
$\zeta/\zeta_{\rm non}$, where $\zeta$ is the bulk viscosity that takes into account all weak 
processes, while $\zeta_{\rm non}$ is an approximate result, see Eq.~(\ref{zeta_non}), in which 
only the nonleptonic processes are included and the semi-leptonic processes are not. When 
the ratio $\zeta/\zeta_{\rm non}$ is substantially larger than 1, it is an indication that the 
semi-leptonic processes play an important role and, thus, cannot be neglected. 

Compared to the normal phase result, the main features of the temperature dependences in  
spin-one color-superconducting phases (see the upper panels in Fig.~\ref{results-set-A}) 
are (i) a smoothed shape of the semi-leptonic ``hump" and (ii) an overall enhancement of 
the bulk viscosity due to color superconductivity for a substantial range of temperatures 
below $T_c$. 

\begin{figure*}[t]
\includegraphics[width=0.48\textwidth]{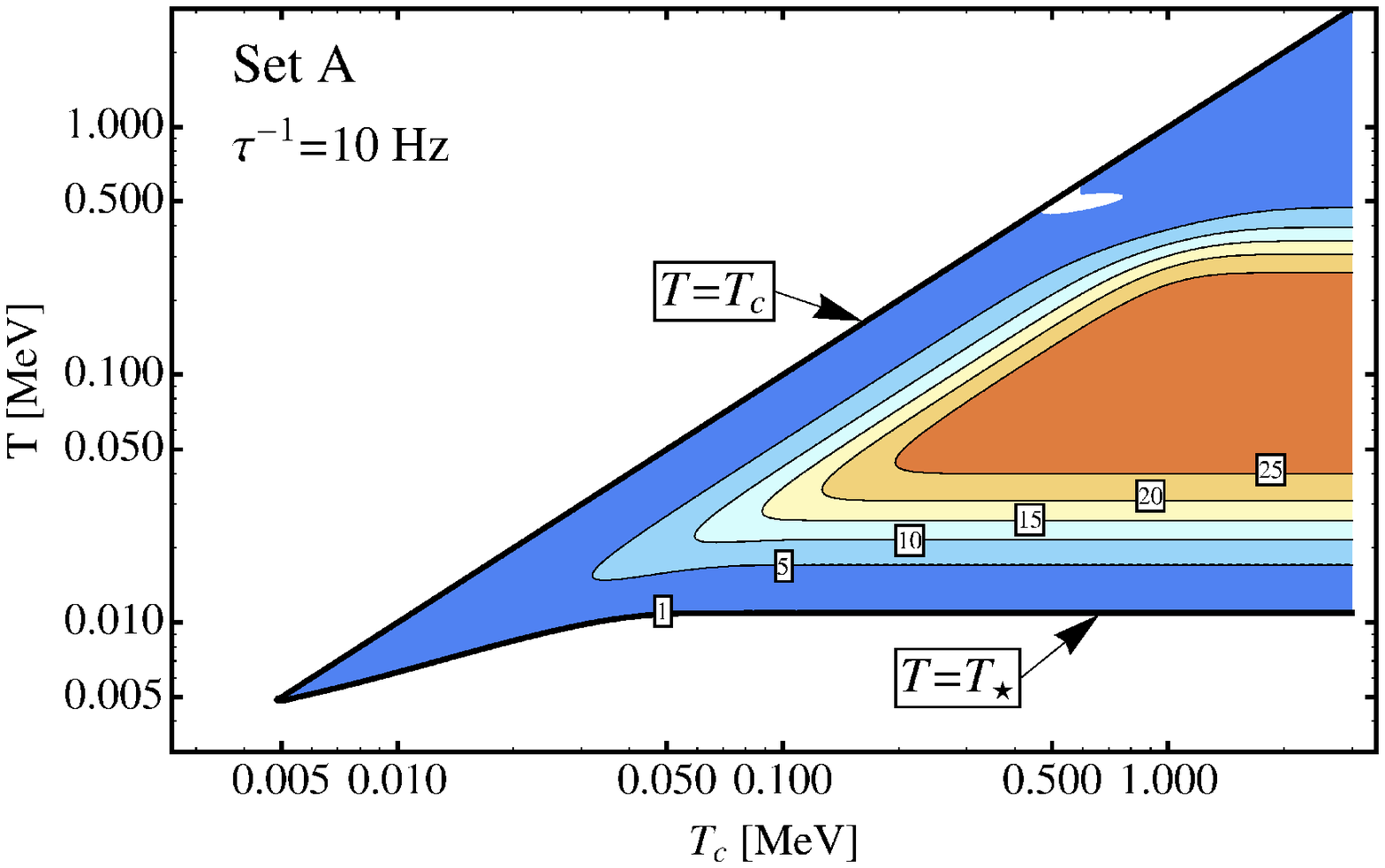}\hspace{0.03\textwidth}
\includegraphics[width=0.48\textwidth]{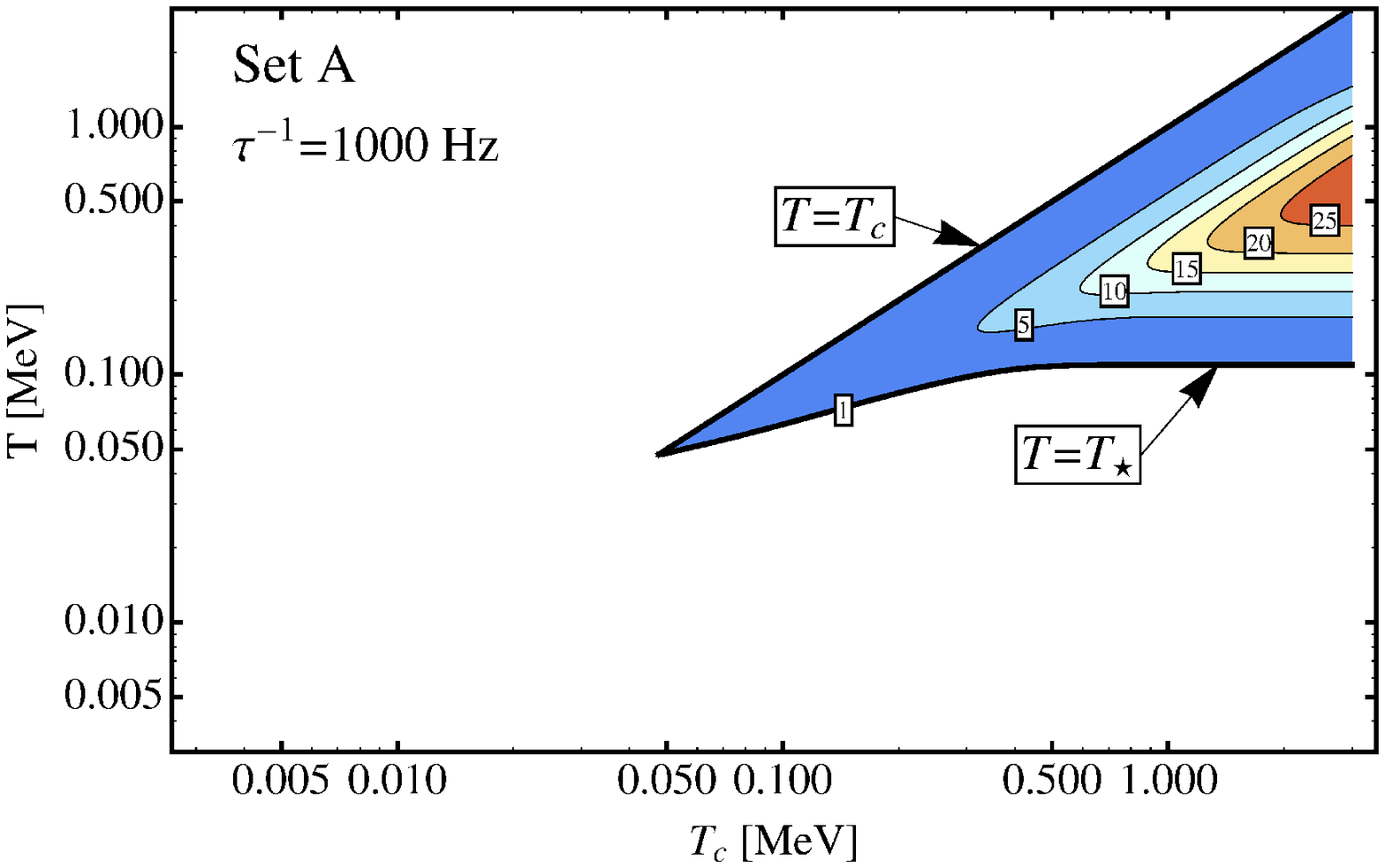}
\caption{(color online). The contour plot of the bulk viscosity enhancement factor due to spin-one 
color superconductivity. The results are for the CSL phase in a model with the parameters in Set A.
The results for two frequencies of the density oscillations are shown: $\tau^{-1}=10~\mbox{Hz}$ (left 
panel) and $\tau^{-1}=1000~\mbox{Hz}$ (right panel). The ratio $\zeta_{\rm CSL}/\zeta_{\rm normal}$  
is larger than $1$ in the colored regions and is equal to or less than $1$ in the white region. The 
contours are labeled by the corresponding values of the ratio $\zeta_{\rm CSL}/\zeta_{\rm normal}$.}
\label{diagram-T-vs-Tc}
\end{figure*} 

As in the case of the normal phase, the semi-leptonic processes are responsible for an 
increase (``hump") of the bulk viscosity in a region of temperatures around $T_{\rm hump}$, 
where 
\begin{subequations}
\begin{eqnarray}
T_{\rm hump}^{\rm (Set~A)}&\simeq& 2.1~\mbox{MeV} \left(\frac{1~\mbox{ms}}{\tau}\right)^{1/4},\\
T_{\rm hump}^{\rm (Set~B)}&\simeq& 1.4~\mbox{MeV} \left(\frac{1~\mbox{ms}}{\tau}\right)^{1/4}
\end{eqnarray}
\label{T_hump}
\end{subequations}
are the approximate positions of the peak of the hump in the normal phase in the case 
of the model parameters in Set A and Set B, respectively. In order to derive these results, 
we used an approximate expression for the bulk viscosity in Eq.~(24) of Ref.~\cite{Sa'd:2007ud},
which is valid when the nonleptonic rate is infinitely large while the semi-leptonic rates are finite.
The maximum of that expression corresponds to $\lambda_2+\lambda_3=n \omega A_1/(B_1A_2-B_2A_1)$, 
whose solution determines an approximate value for $T_{\rm hump}$. Two remarks are in order here: 
(i) the scaling law $T_{\rm hump}\propto 1/\tau^{1/4}$ follows from the power-law temperature 
dependence of the semi-leptonic rates $\lambda_2,\lambda_3\propto T^4$ and (ii) the overall value 
in Eq.~(\ref{T_hump}) is slightly corrected to match the actual numerical results in the case of a
finite nonleptonic rate. 

When $T_c \gtrsim T_{\rm hump}$ the semi-leptonic hump is partially washed out by the 
presence of color superconductivity. This is most clearly seen from the ratio of the bulk 
viscosities $\zeta/\zeta_{\rm non}$ in the lower panels in Fig.~\ref{results-set-A}. While 
the inclusion of the semi-leptonic processes leads to an increase of the viscosity, 
the effect is not as large as in the normal phase. Of course, this conclusion is sensitive 
to the choice of the color-superconducting critical temperature $T_c$. In general, two 
qualitatively different regimes can be realized. When $T_c\lesssim T_{\rm hump}$, the 
hump occurs in the normal phase and, therefore, its shape is almost unaffected by color 
superconductivity. In the opposite case, $T_c\gtrsim T_{\rm hump}$, the effect is 
present and gets stronger as $T_c$ increases relative to $T_{\rm hump}$. 

Now, let us turn to an overall enhancement of the bulk viscosity due to color 
superconductivity below $T_c$. This is observed almost for the whole range of temperatures 
$T_{0,{\rm max}} \lesssim T \leq T_c$, where $T_{0,{\rm max}}$ is the temperature at 
which the bulk viscosity of the normal phase has a global maximum. The value of 
$T_{0,{\rm max}}$ can be easily estimated by considering an approximate expression 
for the bulk viscosity (\ref{zeta_non}) when only the nonleptonic processes are taken 
into account. The maximum of Eq.~(\ref{zeta_non}) corresponds to $\alpha_1=A_1$. 
After solving this for the temperature, we obtain 
\begin{subequations}
\begin{eqnarray}
T_{0,{\rm max}}^{\rm (Set~A)} &\simeq&  47~\mbox{keV} \sqrt{\frac{1~\mbox{ms}}{\tau}},\\
T_{0,{\rm max}}^{\rm (Set~B)}&\simeq&  41~\mbox{keV} \sqrt{\frac{1~\mbox{ms}}{\tau}},
\end{eqnarray}
\label{T0max}
\end{subequations}
where $\tau$ is the period of oscillations measured in milliseconds. Notably, the location
of the maximum is almost the same for both sets of model parameters. Because of the 
superconductivity, the location of the maximum is shifted to a higher temperature, 
$T_{\phi,{\rm max}}\simeq T_{0,{\rm max}}/\sqrt{\cal N}$, where ${\cal N}$ is the same 
parameter that appears in Eq.~(\ref{lambda_1_general}). Taking the shift of the maximum 
into account, we find that the enhancement relative to the normal phase is observed for 
$T_{\rm \star} \leq T \leq T_c$ with $T_{\rm \star} \simeq T_{0,{\rm max}}/{\cal N}^{1/4}$ 
being the point between $T_{0,{\rm max}}$ and $T_{\phi,{\rm max}}$, at which the bulk 
viscosities for the normal and superconducting phases cross. At lower temperatures, 
$T<T_{\rm \star}$, the effect of color superconductivity is opposite: it reduces the bulk 
viscosity. 

\begin{figure*}[t]
\includegraphics[width=0.48\textwidth]{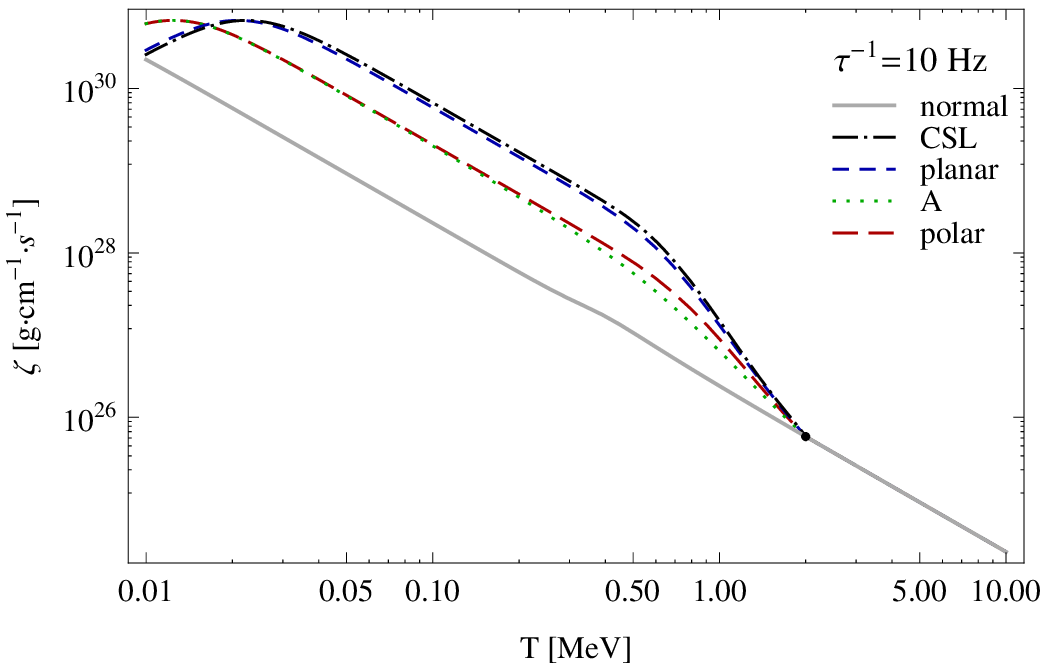}\hspace{0.03\textwidth}
\includegraphics[width=0.48\textwidth]{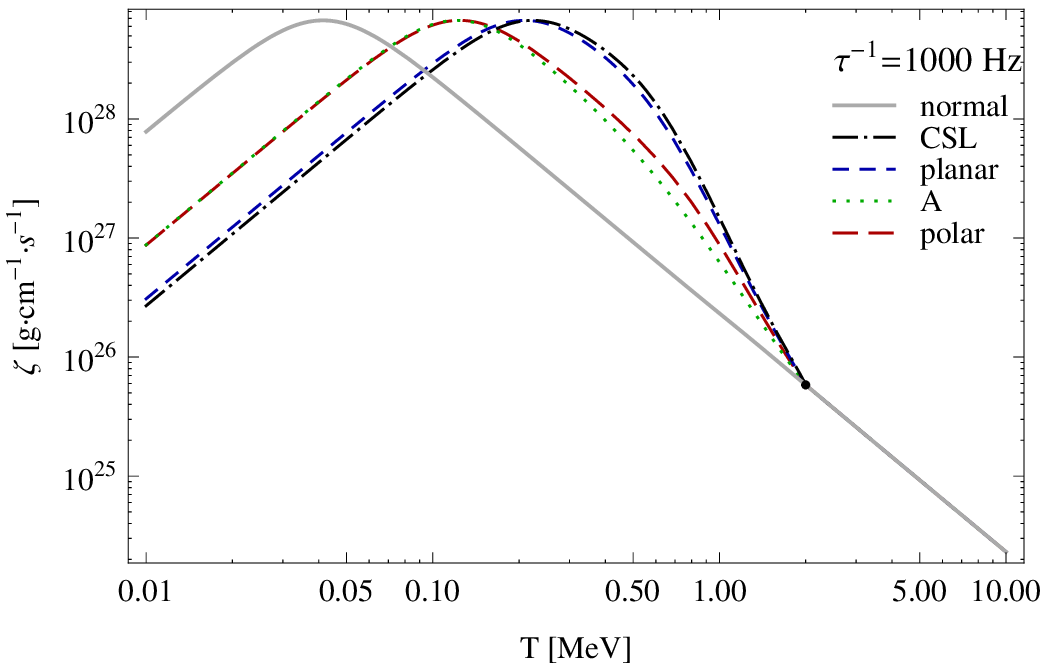}\\
\includegraphics[width=0.48\textwidth]{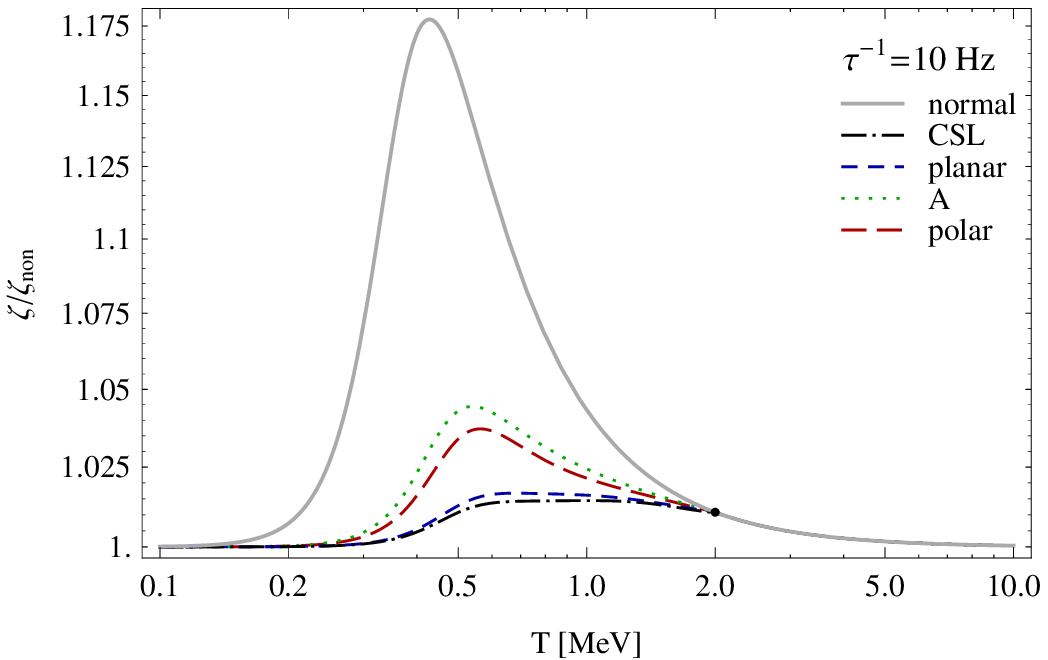}\hspace{0.03\textwidth}
\includegraphics[width=0.48\textwidth]{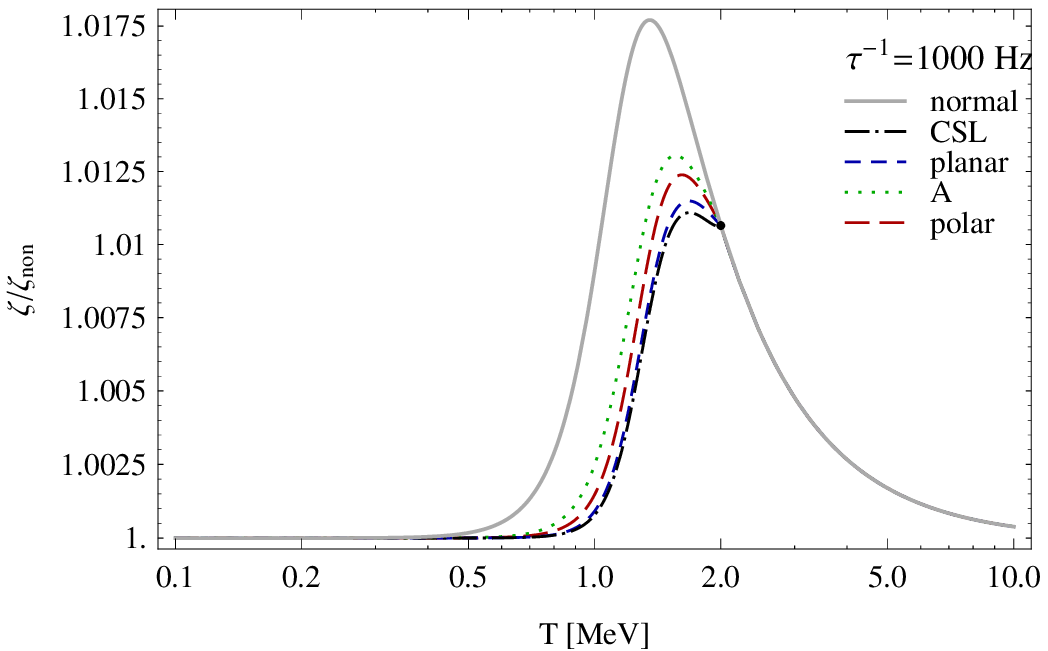}
\caption{(color online). Temperature dependence of bulk viscosity $\zeta$ and the ratio $\zeta/\zeta_{\rm non}$  
for model parameters in Set B and the spin-one color superconducting critical temperature $T_c=2~\mbox{MeV}$. 
The results for two fixed frequencies of the density oscillations are shown.}
\label{results-set-B}
\end{figure*} 
 
The range of temperatures, in which the bulk viscosity increases relative to the normal 
phase of quark matter, depends on the value of the critical temperature $T_c$ and the
frequency of oscillations. While the actual enhancement of the viscosity also depends 
on the specific pattern of spin-one pairing, the qualitative features in all four phases 
studied here are similar. As an example, let us consider the CSL phase in more detail.
In Fig.~\ref{diagram-T-vs-Tc}, we show the contour plot for the bulk viscosity enhancement 
factor due to color superconductivity. The ratio $\zeta_{\rm CSL}/\zeta_{\rm normal}$ 
is larger than $1$ only in the colored regions in Fig.~\ref{diagram-T-vs-Tc}. In white
regions, it is either $1$ (when $T>T_c$) or less than $1$ (otherwise). 

As evident from Fig.~\ref{diagram-T-vs-Tc}, the enhancement of the bulk viscosity by 
spin-one color superconductivity occurs in a rather wide range of temperatures, especially 
when the frequency of density oscillations is not too large and the value of $T_c$ is not 
too small. At $\tau^{-1}=10~\mbox{Hz}$, for example, it extends over an order of magnitude 
or more in temperature, provided $T_c\gtrsim 100~\mbox{keV}$. At $\tau^{-1}=1000~\mbox{Hz}$, 
in contrast, an order of magnitude or wider temperature range for the enhancement is 
seen only if $T_c\gtrsim 1~\mbox{MeV}$. (It should be noted that, in the case 
$\tau^{-1}=10~\mbox{Hz}$ shown in the left panel in Fig.~\ref{diagram-T-vs-Tc}, the 
ratio $\zeta_{\rm CSL}/\zeta_{\rm normal}$ is truly less than $1$ in a small white region 
just below the $T=T_c$ line. This ``abnormality" is due to a subtle interplay between 
the semi-leptonic and nonleptonic processes when the value of $T_c$ is fine-tuned to 
be near $T_{\rm hump}$.)

By ignoring the subtle complications due to the semi-leptonic hump around $T_{\rm hump}$, 
we find that the enhancement of the bulk viscosity in the window of temperatures 
$T_{\rm \star} \leq T \leq T_c$ (as well as the suppression at lower temperatures, 
$T<T_{\rm \star}$) is primarily due to the reduction of the nonleptonic rate $\lambda_1$ 
in color-superconducting phases. At temperatures below $T_c$, when all gapped 
quasiparticles effectively cease to contribute, the corresponding reduction factor 
for the rate is approximately given by the value of ${\cal N}$. This means that the 
enhancement factor for the viscosity approaches its inverse value, ${\cal N}^{-1}$. By 
making use of the numerical results for ${\cal N}$, we find that the enhancement factor 
for the bulk viscosity reaches up to about $9$ in the $A$- and polar phases, $25$ in the 
planar phase and $29$ in the CSL phase. (The suppression factors at $T<T_{\rm \star}$ 
approach the same values.) In the region of the hump, of course, the behavior is 
more complicated, but the overall effect of superconductivity is still mainly to increase 
the bulk viscosity. 

The numerical results in the case of the model parameters in Set B are shown in 
Fig.~\ref{results-set-B}. The qualitative features are similar to those obtained for Set A. 
However, the effect of the semi-leptonic processes is less pronounced: the corresponding 
hump is almost non-existent and the ratio $\zeta/\zeta_{\rm non}$ does not much deviate 
from 1. At the same time, the effect of color superconductivity is very well pronounced. 
Compared to the normal phase result, an enhancement of the bulk viscosity by a factor of 
about ${\cal N}^{-1}$ is seen in a relatively wide window of temperatures from $T_{\rm \star}$ 
to $T_c$. 

\section{Discussion}
\label{Discussion}

In this study, we calculated the bulk viscosity in spin-one color-superconducting strange 
quark matter by carefully taking into account the interplay between the nonleptonic and 
semi-leptonic week processes. 

As expected, the nonleptonic processes give the dominate 
contribution to the viscosity in a wide range of parameters. Yet, as in the normal phase 
\cite{Sa'd:2007ud}, the semi-leptonic processes may also lead to a substantial correction 
in a window of temperatures around $T_{\rm hump}$, see Eq.~(\ref{T_hump}). The value of 
$T_{\rm hump}$ scales as $1/\tau^{1/4}$ and happens to be of order $1~\mbox{MeV}$ for 
millisecond pulsars. The size and the relative importance of the hump can be conveniently
measured by the ratio $\zeta/\zeta_{\rm non}$ when it is noticeably larger than $1$. 
For millisecond pulsars, however, this ratio remains close to $1$. The effect is more 
pronounced when the period is a few orders of magnitude longer. We also find that the 
corresponding hump in the temperature dependence of the bulk viscosity of color 
superconductors is partially washed out compared to the normal phase. The higher 
is $T_c$ relative to $T_{\rm hump}$, the larger wash out of the hump is seen. At sufficiently 
low $T_c$, i.e., $T_c \lesssim T_{\rm hump}$, the hump occurs in the normal phase and, 
therefore, its shape is unaffected by color superconductivity. 

If the critical temperature of the spin-one color-superconducting phase transition $T_c$ is 
larger than $T_{0,{\rm max}}$, see Eq.~(\ref{T0max}), the main effect of color superconductivity 
is an overall increase of the bulk viscosity in a range of subcritical temperatures, 
$T_{\rm \star} \leq T \leq T_c$, see Fig.~\ref{diagram-T-vs-Tc}. The corresponding range 
of temperatures widens with increasing the value of $T_c$ and with decreasing the frequency 
of oscillations. The increase of the viscosity is primarily due to the suppression 
of the nonleptonic rate by color superconductivity. At almost all temperatures below 
$T_c$, the rate is dominated by the ungapped quasiparticles, whose relative contribution 
is scaled by the factor ${\cal N}$ with respect to the normal phase (note that ${\cal N}<1$). 
It is the inverse value ${\cal N}^{-1}$ that determines the maximal enhancement 
of the bulk viscosity at subcritical temperatures. The corresponding enhancement 
factor is equal to $9$ in the $A$- and polar phases, about $25$ in the planar phase 
and about $29$ in the CSL phase.  (At temperatures below $T_{\rm \star}$, color 
superconductivity leads to a suppression of the bulk viscosity, and the maximal 
suppression will approach the same value of ${\cal N}^{-1}$.) 

In relation to this result, it might be appropriate to note that a similar enhancement 
mechanism was previously observed for spin-zero color superconductors 
\cite{Alford:2006gy}. A special feature of spin-one color superconductivity is 
that the maximum enhancement factor can be much larger.

In our analysis, we utilized the same spin-one pairing pattern as in 
Refs.~\cite{Schafer:2000tw,Schmitt:2002sc,Schmitt:2004et}. In the case of zero quark 
masses, the main signature of the corresponding phases is the presence of ungapped 
quasiparticles. When quarks have small masses, the gaps of the corresponding modes 
are of order $\phi m/\mu$. These may be still too small to significantly affect our main 
results. However, if the spin-one gaps are larger, as some studies suggest 
\cite{Marhauser:2006hy},  the suppression of the nonleptonic rates and, therefore, the 
enhancement of the bulk viscosity in color superconducting matter may turn out to be 
even stronger. 

In application to compact stars, we may speculate that the transition to a spin-one color 
superconducting phase in a stellar core can have a stabilizing effect against the r-modes 
driven by the gravitational radiation \cite{Andersson:1997xt}. If the critical temperature 
of the corresponding phase transition is on the order of or above $1~\mbox{MeV}$, 
the corresponding dynamics can affect even relatively young stars. The study of the 
actual quantitative effect that spin-one color superconductivity has on the reduction of 
the instability window in the pulsar frequency and temperature plane can be done along 
the lines of Refs.~\cite{Madsen:1999ci,Andersson:2001ev,Jaikumar:2008kh}. However, 
this problem is beyond the scope of the present paper.

\acknowledgments 
The authors would like thank Mark Alford, Thomas Sch\"afer, Andreas Schmitt and Kai Schwenzer 
for useful comments.  The work of X.W. is also supported in part by the Arizona State University Graduate 
Fellowship. The work of I.A.S. is supported in part by the start-up funds from the Arizona State 
University and by the U.S. National Science Foundation under Grant No. PHY-0969844.

\appendix

\section{$\lambda$-rates of semi-leptonic (Urca) processes}
\label{Urca-rates}

The rates of the semi-leptonic processes in spin-one color superconducting quark matter
were calculated in Ref.~\cite{Sa'd:2006qv}. The general expression for the rate takes the 
following form:
\begin{equation}
\lambda_{i} = \lambda_{i}^{(0)} 
\left[\frac13 +\frac23 H\left(\frac{\phi}{T}\right)\right]
\quad\mbox{for}\quad i=2,3,
\end{equation}
where $ \lambda_{i}^{(0)} $ is the corresponding rate in the normal phase of quark matter
and $H\left({\phi}/{T}\right)$ is a phase-specific suppression factor. By construction, it 
satisfies the constraint $H(0)=1$, which corresponds to the case of the normal phase. We used 
the numerical data of Ref.~\cite{Sa'd:2006qv} to obtain the following fits for the suppression 
factors as functions of the dimensionless ratio $\varphi\equiv {\phi}/{T}$ in the four spin-one 
color superconducting phases of quark matter:
\begin{equation}
H^{A}\left(\varphi\right) 
= \frac{ a_1\varphi^4 +b_1 \varphi^3 + c_1 \varphi^2 + d_1}
{\varphi^5 + e_1 \varphi^3 + f_1 \varphi^2 + d_1} ,
\end{equation}
where $a_1 = 1.069$, $b_1=- 0.2187$, $c_1 = 3.666$, $d_1= 21.50$, $e_1 = 1.333$ 
and $f_1 = 9.349$,
\begin{equation}
H^{\rm polar}\left(\varphi\right) 
= \frac{a_2 \varphi^3 + b_2 \varphi^2 + c_2}
{\varphi^5 + d_2 \varphi^4 + e_2 \varphi^3 + f_2 \varphi^2 + c_2},
\end{equation}
where $a_2 = \pi$, $b_2=21.94$, $c_2 = 1386$, $d_2= 6.994$, $e_2 = 11.20$ and $f_2 = 214.0$,
\begin{equation}
H^{\rm planar}\left(\varphi\right) 
= \frac{a_3 \varphi^{3.5}+ b_3 \varphi^3+ c_3\varphi^2+d_3(1+\varphi)}
{\varphi^3+ e_3\varphi^2 +d_3} e^{-\varphi},
\end{equation}
where $a_3 = 0.917$, $b_3 = 0.456$, $c_3 = 11.69$, $d_3 = 34.0$ and $e_3 = 4.221$,
\begin{equation}
H^{\rm CSL}\left(\varphi\right) 
= \frac{a_4\varphi^4+b_4\varphi^3+c_4\varphi^2+d_4(1+\sqrt{2}\varphi)}
{\varphi^3+e_4\varphi^2+d_4} e^{-\sqrt{2}\varphi} ,
\end{equation}
where $a_4 = 1.034$, $b_4 = 1.001 $, $c_4 = 9.735$, $d_4 = 13.81 $ and $e_4 = 1.684$.

\section{$\lambda$-rates of nonleptonic processes}
\label{nonleptonic-rates}

The $\lambda$-rate of the nonleptonic processes in spin-one color superconducting quark 
matter was calculated in Ref.~\cite{Wang:2009if}. The general expression for the rate 
takes the following form:
\begin{equation}
\lambda_{1} = \lambda_{1}^{(0)} 
\left[{\cal N} + (1-{\cal N})\tilde{H}\left(\frac{\phi}{T}\right)\right] ,
\end{equation}
where $\lambda_{1}^{(0)} $ is the corresponding rate in the normal phase of quark matter, 
${\cal N}$ is a constant that determines the relative contribution of ungapped quasiparticles 
to the rate, and $\tilde{H}\left({\phi}/{T}\right)$ is a phase-specific suppression factor 
due to gapped quasiparticles. The normal phase corresponds to $\phi=0$, in which case there is 
no suppression and $\tilde{H}(0)=1$. The value of ${\cal N}$ for each phase reads
\begin{subequations}
\begin{eqnarray}
{\cal N}^{A} &=& \frac19,\\
{\cal N}^{\rm polar} &=& \frac19 ,\\
{\cal N}^{\rm planar} &\approx & 0.0393 ,\\
{\cal N}^{\rm CSL} &=& \frac{928}{27027}.
\end{eqnarray}
\end{subequations}
For this study we used the numerical data of Ref.~\cite{Wang:2009if} to obtain the 
following fits for the suppression factors as functions of the dimensionless ratio 
$\varphi\equiv {\phi}/{T}$:
\begin{equation}
\tilde{H}^{A}\left(\varphi\right) 
= \frac{\alpha_1 \varphi^2+ \beta_1}
{\varphi^3+  \gamma_1\varphi^2+\beta_1},
\end{equation}
where $\alpha_1 = 0.1247 $, $\beta_1 = 12.60 $ and $\gamma_1 = 5.042 $, 
\begin{equation}
\tilde{H}^{\rm polar}\left(\varphi\right) 
= \frac{\alpha_2 \varphi^2+ \beta_2}
{\varphi^4+\gamma_2 \varphi^2+\beta_2},
\end{equation}
where $\alpha_2 = 0.0271 $, $\beta_2 = 65.45 $ and $\gamma_2 = 13.35$, 
\begin{equation}
\tilde{H}^{\rm planar}\left(\varphi\right) 
= \frac{\alpha_3\varphi^{4}+ \beta_3\varphi^3+ \gamma_3\varphi^2+\delta_3(1+\varphi)}
{\varphi^2+ \delta_3} e^{-\varphi},
\end{equation}
where $\alpha_3 = 0.0717$, $\beta_3 = -0.2663$, $\gamma_3 = 1.108$ and $\delta_3 = 4.561$,
\begin{equation}
\tilde{H}^{\rm CSL}\left(\varphi\right) 
= \frac{\alpha_4 \varphi^4+\beta_4\varphi^2+\gamma_4(1+\sqrt{2}\varphi)}
{\varphi^3+\delta_4\varphi^2+\gamma_4} e^{-\sqrt{2}\varphi} ,
\end{equation}
where $\alpha_4 = 0.6981 $, $\beta_4 = -2.045$, $\gamma_4 = 4.482$ and $\delta_4 = -1.217$.

\end{document}